\documentclass[prl,footinbib,amsmath,amssymb,showpacs,twocolumn,superscriptaddress]{revtex4}
\usepackage{graphicx}
\usepackage{graphics,color}
\usepackage{ulem}

\newcommand{\bs}{\;\;\;\;\;}
\newcommand{\sms}{\;\;}
\newcommand{\ve}{\mathbf}

\begin{document}

\title{Optical manipulation of edge state transport in HgTe quantum wells in the quantum hall regime}
\author{M. J. Schmidt}
\affiliation{Department of Physics, University of Basel, Klingelbergstrasse 82, 4056 Basel, Switzerland}
\author{E. G. Novik}
\affiliation{Physikalisches Institut (EP3), University of W\"urzburg, 97074 W\"urzburg,
Germany}
\author{M. Kindermann}
\affiliation{School of Physics, Georgia Institute of Technology, Atlanta, Georgia 30332, USA}
\author{B. Trauzettel}
\affiliation{Institute for Theoretical Physics and Astrophysics, University of W\"urzburg, 97074 W\"urzburg, Germany}
\date{\today}

\begin{abstract}
We investigate an effective low energy theory of HgTe quantum wells near their mass inversion thickness in a perpendicular magnetic field. By comparison of the effective band structure with a more elaborated and well-established model, the parameter regime and the validity of the effective model is scrutinized. Optical transitions in HgTe quantum wells are analyzed. We find selection rules which we functionalize to optically manipulate edge state transport. Qualitatively, our findings equally apply to optical edge current manipulation in graphene.
\end{abstract}

 \maketitle

Low dimensional quantum systems with distinct topological properties attract a lot of interest, not only because of their possible applications in topological quantum computation \cite{kitaev}, but also because they constitute a versatile playground for studying solid state realizations of exotic phases \cite{wen_book}. It has recently been realized \cite{effective_model} that HgTe quantum wells (QWs) exhibit very rich low energy properties such as the quantum spin Hall effect and topological phase transitions. Notably, these extraordinary physical properties are not only theoretical predictions but many of them have already been experimentally confirmed in HgTe nanodevices \cite{koenig,alena_paper,hartmut_AC_effect}. The remarkable tunability of parameters like, for instance, the Rashba spin orbit coupling (SOC) strength \cite{hartmut_AC_effect} makes HgTe QWs especially suitable for spintronics applications. Interestingly, the low-energy properties of electrons in HgTe QWs near the so-called inversion point (which is related to the thickness of the HgTe layer) can be well described by the Dirac equation similar to the low-energy properties of electrons in graphene \cite{Geim2007}. However, the electronic spectrum of HgTe QWs is much richer than that of graphene and further theoretical as well as experimental research is needed to fully characterize it.

In this Rapid Communication, we first compare an effective model for HgTe QWs \cite{effective_model} (in the presence of a perpendicular magnetic field) with a more elaborated 8-band Kane model \cite{alena_paper,alena_paper2}. This allows us to identify the experimentally relevant parameter regime for the effective model. After the important energy scales are identified, we can analyse the optical transitions between Landau levels and, even more interestingly, between (magnetic-field induced) edge states. We show that this gives rise to a new possibility of edge current reversion by photons. This effect is not unique to HgTe QWs but can happen in all systems where optical transitions between electron-like and hole-like edge states are allowed which is, for instance, the case in graphene.

\begin{figure}[here]
\centering
\includegraphics[width=240pt]{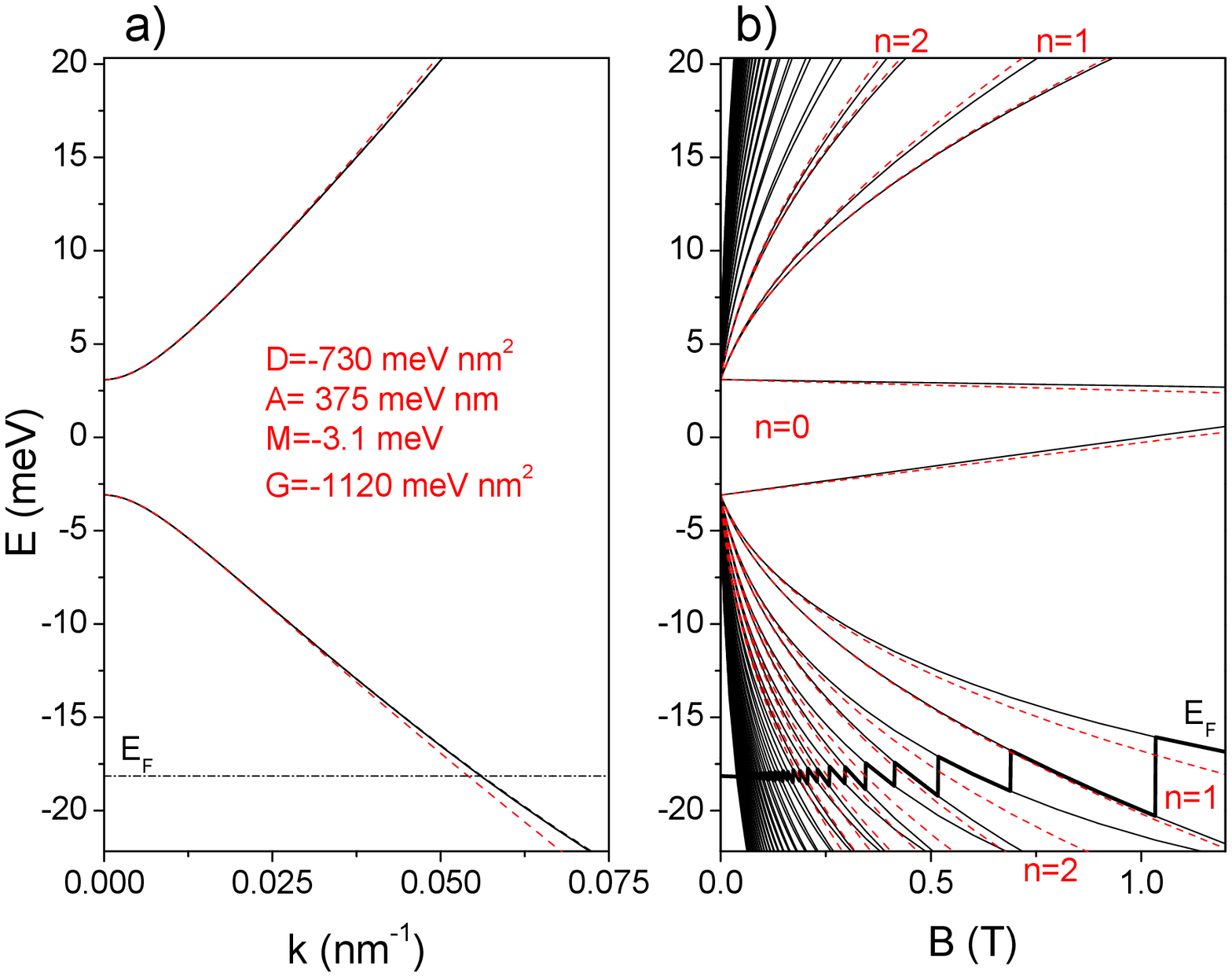}
\caption{(Color online) (a) Energy spectra of the symmetrically doped QW at zero magnetic field for a hole density of $0.5 \cdot 10^{11} {\rm cm}^{-2}$ and a well thickness $d=6.5{\rm nm}$. Solid (black) lines correspond to the 8-band Kane model, and dashed red (gray) lines show the energy subbands given by the effective model with the optimal parameters $D=-730 {\rm meV\, nm}^2,\, A=375 {\rm meV\, nm}, \, M=-3.1 {\rm  meV}, \, G=-1.120 {\rm eV\, nm}^2$. (b) Landau levels in the effective model (dashed, red lines) and in the Kane model (solid, black lines), using the same parameters as in (a).}
\label{comparison_figure}
\end{figure}

The bulk band structure of narrow HgTe QWs has been extensively investigated before \cite{alena_paper,alena_paper2}. It was found that near the mass inversion thickness $d_c\simeq6.3{\rm nm}$, the electronic properties close to the $\Gamma$ point are well approximated within an effective theory \cite{effective_model}, defined by the $\ve k$-diagonal Hamiltonian \footnote{We set $\hbar,e,c=1$ in the following.}
\begin{equation}
H=\left(\begin{matrix}h(\ve k) & 0 \\ 0 & h^*(-\ve k) \end{matrix}\right),\bs h(\ve k) = \epsilon(\ve k) + d_a(\ve k)\sigma^a \label{effective_model_hamiltonian}
\end{equation}
with $\sigma^a$ the Pauli matrices, $\epsilon(\ve k) = - D k^2$, $\ve d(\ve k) = (A k_x, -A k_y, M-G k^2)$ and $\ve k$ the crystal momentum. $A,D,M$ and $G$ are parameters of the effective model (see Fig. \ref{comparison_figure}). In this model,
Rashba and Dresselhaus spin orbit coupling (SOC) are not explicitly taken into account. Rashba SOC could be treated easily by perturbation theory and can, in principle, be tuned to zero in the experiments. Therefore, we neglect it here. Further, the Dresselhaus terms are known to be negligibly small \cite{koenig}.
The basis states of the model in Eq.~(\ref{effective_model_hamiltonian}) are $\{\left|E+\right>,\left|H+\right>,\left|E-\right>,\left|H-\right>\}$ where $E$ ($H$) refers to the subband which is predominantly derived from the conduction (valence) band, and $\pm$ refers to degenerate Kramers partners \cite{alena_paper}.

The presence of a time reversal symmetry breaking magnetic field perpendicular to the well is described by a corresponding vector potential - we use the Landau gauge here. After the usual transformations, it is found that the Hamiltonian which describes the Landau levels is obtained by the replacement
\begin{equation}
h(\ve k) \rightarrow h_+,\bs h^*(-\ve k) \rightarrow h_-
\end{equation}
in Eq. (\ref{effective_model_hamiltonian}) with $h_{\pm} = h_{HO} + h_{JC}^\pm$. The harmonic oscillator part of the Hamiltonian $h_{HO} = -2 DB(a^\dagger a + \frac12) +\left[M-2BG(a^\dagger a+\frac12)\right]\sigma^3$ is diagonal in the electron-hole space as well as in the Kramers space, while the Jaynes-Cummings terms $ h_{JC}^+ = -\sqrt{2B}A(a^\dagger \sigma^+ + a \sigma^-)$ and $h_{JC}^- = +\sqrt{2B}A(a \sigma^+ + a^\dagger \sigma^-)$ couple different harmonic oscillator levels and lift the $\pm$ degeneracy. $a$ and $a^\dagger$ are bosonic operators, i.e. $[a,a^\dagger]=1$. Thus, the $2\times 2$ blocks $h_\pm$ are easily diagonalized by introducing $\left|1_n\right>_+ = \left|n\right>\otimes\left|E+\right>$, $\left|2_n\right>_+ = \left|n-1\right>\otimes\left|H+\right>$ and $\left|1_n\right>_- = \left|n\right>\otimes\left|H-\right>$, $\left|2_n\right>_- = \left|n-1\right>\otimes\left|E-\right>$, respectively, where $\left|n\right>$ are harmonic oscillator states corresponding to the Landau level $n$.


In order to find the proper parameters for the effective model (\ref{effective_model_hamiltonian}) we perform extensive calculations within a well established 8-band $\ve k\cdot \ve p$ approach for the HgTe QW \cite{alena_paper,alena_paper2}. The parameters which enter this numerical calculation are the well known band structure parameters for HgTe and CdTe, the width of the QW, the strength of the magnetic field $B$ and the charge density in the 2DEG. The parameters $A,M,D$ and $G$ of the effective model are determined in the limits $|\ve k|\rightarrow0$ and $B\rightarrow0$. The results are shown in Fig. \ref{comparison_figure}, correcting and extending previous estimates made in Refs. \cite{effective_model,koenig}. As expected, the effective model is best near the $\Gamma$-point. In a finite magnetic field perpendicular to the 2DEG plane, the low energy states acquire additional components from higher $|\ve k|$ states. Thus, the effective model description of Landau levels becomes worse at higher magnetic fields. For $B \lesssim 1T$, however, the effective model is still reasonably accurate.

\begin{figure}[here]
\centering
\includegraphics[width=250pt]{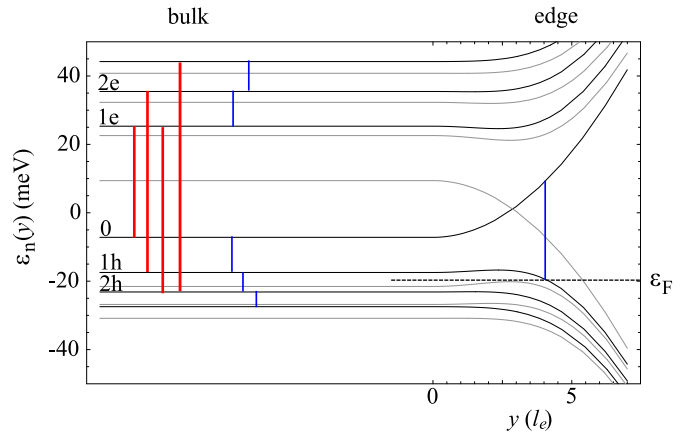}
\caption{(Color online) Landau level dispersion and optical transitions for $h_+$ for $B=1T$ in bulk and near an edge. The electron- (hole-) like levels are labeled by $n\rm e$ ($n\rm h$) \cite{e_h_like_footnote}. The thick (red) lines are the interband transitions. The thin (blue) lines are intraband transitions of smaller energy in the bulk. The thin (blue) line in the edge region represents the transition which we choose (see text). The energy levels of $h_-$ are drawn in light gray.}
\label{edge_states_figure}
\end{figure}

In the following, we focus on $h_+$ since there is no coupling between $h_+$ states and $h_-$ states. The energy spectrum of the $h_+$ eigenstates is plotted in Fig. \ref{edge_states_figure}. The treatment of the $h_-$ block works analogously. For completeness, these $h_-$ energy states are drawn in light gray in Fig. \ref{edge_states_figure}.

Since we aim at an investigation of finite 2D structures, an important issue is the modelling of the edges and the corresponding edge states.
We introduce an edge into our Hamiltonian by a mass term $M\rightarrow M + V_{edge}(y)$ that varies slowly on the scale of the typical extent of a Landau wave function perpendicular to the edge ($y$-direction). In the then justified adiabatic approximation one obtains an extra contribution $V_{edge}(k/B)\sigma^3$ to the electron energies, where $k$ is the crystal momentum parallel to the edge. The resulting energy diagram of the electron states near an edge with quadratic confinement potential $V_{edge}(y)=(y/l_e)^2 \rm meV$ is shown in Fig. \ref{edge_states_figure}. $l_e$ is the typical length scale of the variation of the confinement potential. We further assume that the two edges at opposite sides are sufficiently away from each other such that the finite size effects of overlapping edge states, recently analyzed in Ref.~\cite{Zhou08}, do not matter.

Irradiation by a classical electromagnetic field is described by   a time dependent vector potential $\ve A_1(\ve r,t) = 2 |A_1| \hat\epsilon \cos(\omega/c\, \hat n\cdot\ve r - \omega t)$.  We choose $\omega$ in the far infrared (FIR) regime, corresponding to the typical excitation energies in our system.
Consider linearly polarized light ($\hat\epsilon = \hat e_y$) shining on the sample from the $+z$ direction ($\hat n = -\hat e_z$), so that we can write in the Landau gauge
\begin{equation}
k_x \rightarrow p_x -B y,\bs k_y\rightarrow p_y+2|A_1|\cos(-\omega t).
\end{equation}
To linear order in the radiation field, two new terms appear in the Hamiltonian $h_+ \rightarrow h_+ + h^{\omega1}_+ + h^{\omega2}_+$ with
\begin{align}
h^{\omega1}_+ &= - 2i |A_1| \cos(\omega t) \sqrt{2B}(a^\dagger - a) \left[D I_{2\times2} + G \sigma^3\right] \\
h^{\omega2}_+ &=  - 2|A_1|A \cos(\omega t)\, \sigma^2.
\end{align}
For typical parameters as taken from Fig. \ref{comparison_figure}, $h^{\omega1}_+$ is one order of magnitude smaller than $h^{\omega2}_+$ and thus, we neglect it \footnote{$h^{\omega1}_+$ does not change the selection rules, only the oscillator strengths.}. By employing a rotating wave approximation we find transitions between $\left|\psi^+_i(n)\right>$ and $\left|\psi^+_j(n\pm1)\right>$ where $\left|\psi_i^+(n)\right>,\; n=1,...,\infty,\,i={\rm e,h}$ are the eigenstates of $h_+$ (see Fig. \ref{edge_states_figure}). In addition, there is a nonzero optical matrix element between the zero mode and the electron- and hole-like states of the first Landau level. This situation is very similar to the graphene case \cite{graphene_optical_transition}. The presence of an edge which is modelled by a spatially varying mass term does not change the selection rules $\Delta n = \pm 1$. Also a more elaborate calculation within the 8 band Kane model leads qualitatively to the same optical transitions.

We now turn to the question how  irradiation of a classical light field affects the charge transport through the edge states. Consider one single pair of edge states, namely the $1$h and the zero mode of the $h_+$ sector (see Fig. \ref{edge_states_figure}). Further assume that the Fermi energy is tuned such that it crosses the topmost counterclockwise moving edge state $1$h  \footnote{The counterclockwise (clockwise) moving edge states are those with a negative (positive) dispersion as the edge is approached.}.  One expects that the irradiation of a short, strong FIR pulse, tuned to the transition energy, is able to scatter an electron of the counterclockwise moving edge state (1h) into the clockwise moving edge state (0) with a higher energy and reversed direction of motion. Note that the crystal momentum of the scattered electron stays constant during this process. Counterintuitively, in this system light may thus backscatter electrons  by reversing their group velocity at constant momentum through transitions  from a hole-like band into an electron-like band \cite{e_h_like_footnote}.


A more realistic scenario is the continuous illumination of the whole QW by light, tuned to the selected transition. The relevant properties of an edge are its length $L$, the radiation strength profile $A_1(x)$ as a function of the edge coordinate $x$ and its Fermi energy $\varepsilon_{\rm F}$, defined by the electrochemical potential of the upstream reservoir. In the following, we assume that the frequency of the radiation is tuned to resonance with the transition indicated in Fig.\ \ref{edge_states_figure}. More precisely, we assume that the energetic difference $\Delta \varepsilon$ between the state in the $1$h mode with the Fermi energy and the zero mode state with identical momentum equals $\omega$, the radiation frequency. We focus on this pair of states and neglect all other edge states as they are all highly off-resonant. We also assume a vanishing laser line width. The transport through the edge is then described by a $2\times 2$ transfer matrix $T_E(x,x')$ \cite{mello-kumar} which satisfies the  equation
\begin{equation}
\left[E - H(-i\partial_x)\right] T_E(x,x')=0, \label{defining_equation}
\end{equation}
where $H(k)$ is well approximated by
\begin{equation}
H(k) \simeq \uuline v\,k + Q(x) \sigma^2,\bs \uuline v=\left(\begin{matrix} -v_1 & 0\\0 & v_2 \end{matrix}\right).
\end{equation}
Here, $Q(x)=\gamma A |A_1(x)|$ characterizes the $x$-dependent intensity of the FIR source and  $E$ measures the energy of the scattering states relative to the Fermi energy $\varepsilon_{\rm F}$. $\gamma \lesssim 1$ is a nonuniversal number which depends on the model parameters. We henceforth absorb $\gamma$ into $A_1$. The solution of (\ref{defining_equation}) with the initial condition $T_E(0,0)=1$ yields
\begin{equation}
M_E = T_E(0,L) = T_x\exp\left(-i \int_0^L dx \,\uuline v^{-1} ( E - Q(x)\sigma^2)\right).
\end{equation}
The symbol $T_x$ indicates a spatial ordering of operators, in analogy to the time ordering operator in the quantum mechanical time evolution operator. $v_1$ and $v_2$ are the absolute velocities of the clockwise mover and the counterclockwise mover, respectively. In the linear response regime at zero temperature, when the energy $E$ of the relevant edge state is $E = 0$, we find an exponential suppression of the propagation through the edge. The transmission amplitude at $E=0$ reads
\begin{equation}
t_0(a_1) = \frac1{(M_0)_{22}} = {\rm sech}\left( a_1 \frac{A}{\sqrt{v_1v_2}}\right), \label{transmission_amplitude}
\end{equation}
where $a_1 = \int_0^L dx\, A_1(x)$ is the  FIR intensity integrated over the length of the edge.

\begin{figure}[here]
\centering
\includegraphics[width=210pt]{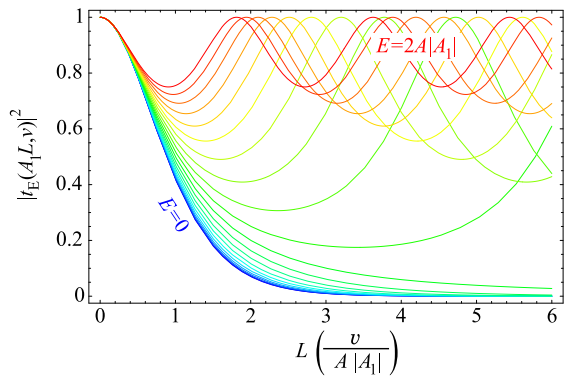}
\caption{(color online) Transmission coefficient $\left|t_E(A_1\,L,v)\right|^2$ away from the linear response regime for counterpropagating modes with identical absolute group velocities $v_1=v_2=v$. $A_1(x) = A_1[\Theta(L-x)-\Theta(-x)]$ is assumed. The energy of the incident electron is varied from the linear response regime $E\simeq 0$ (blue lines) to the regime where the laser is not strong enough to suppress the propagation (red lines).}
\label{transmission_figure}
\end{figure}

The off-resonant edge transmission, $E>0$, is shown in Fig. \ref{transmission_figure} for a steplike radiation intensity profile $A_1(x)=A_1[\Theta(L-x)-\Theta(-x)]$ (where $\Theta(x)$ is the Heaviside function) and $v_1=v_2$. \footnote{A similar radiation-dependence of the transmission coefficient also applies to the case where the whole sample is uniformly illuminated (see below).} As expected, the transport ceases to be exponentially suppressed  at $|E|> A|A_1|$, where  $A|A_1|\sim 10 \mu\rm eV$ for typical parameter values. Transitions between pairs of states with energy difference $\Delta \epsilon$ thus do  not block the  current  if $(\Delta \epsilon - \omega)/A|A_1|\gg 1$. This is typically the case for all pairs of states at equal momentum but the one that  the radiation frequency $\omega$ is tuned to.

We now turn to a discussion of two possible experimental realizations of optical manipulation of edge state transport. For this, we assume that the whole sample is uniformly illuminated by a laser of constant intensity   $|A_1|$, such that $a_1=L|A_1|$.

\begin{figure}[here]
\centering
\includegraphics[width=150pt]{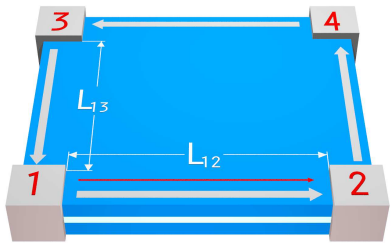}
\caption{(color online) Four terminal setup for edge state experiments. The thick light gray arrows represent the counterclockwise moving edge states with energies below the Fermi energy  $\varepsilon_{\rm F}$. The thin (red) arrow represents the net transport mode with energy larger than $\varepsilon_{\rm F}$ due to a slightly elevated electrochemical potential $\mu_1> \mu_i,\sms i=2,3,4$.}
\label{device_figure}
\end{figure}

{\it Experimental setup I.} Consider a quadratic structure of a HgTe quantum well with four contacts at the corners (see Fig. \ref{device_figure}). The electrochemical potentials of the contacts $\mu_i$ are tuned such that all $\mu_i$ correspond to energies between the bulk Landau levels $1h$ and $2h$ (see Fig. \ref{edge_states_figure}). Within these bounds, assume   that $\mu_1 > \mu_2 = \mu_3=\mu_4$, the difference of $\mu_1$ and $\mu_2$ being sufficiently small to be in the linear response regime. Furthermore we assume equal side lengths $L_{12}=L_{13}$.  Because of $\mu_1>\mu_2$, a net current $I_{12}\propto \mu_1 -\mu_2$ due to a counterclockwise moving transport mode flows from contact 1 to contact 2. Irradiation of laser light, tuned to the transition energy indicated in Fig.\ \ref{edge_states_figure}, results in a suppression of $I_{12}$  by scattering the counterclockwise movers into clockwise moving edge states which exit back into contact 1, as discussed above. The photons of the FIR source supply the required energy for this scattering process.  According to Eq.\  (\ref{transmission_amplitude})   the current $I_{12}$  will be exponentially suppressed
\begin{equation} \label{I12}
I_{12} \sim {\rm sech}^2\left(|A_1| A \frac{L_{12}}v\right)
\end{equation}
in the zero temperature,  linear response regime. Note that Eq.\  (\ref{transmission_amplitude})   has been derived for a laser profile that is focused onto a finite segment of the edge. One, however, expects qualitatively the same behavior also for uniform illumination, which irradiates not only the edge, but also the reservoirs. This is because  the origin of the effect, the opening of a transport gap through coupling of the modes $1$h and $0$, is present in  both scenarios.
For $5\mu \rm m$ of illuminated edge, $v\simeq 10^5 \rm \frac ms$ and a FIR power of 4mW focused onto an area of $1\rm mm^2$ the parameter $ |A_1| A  L_{12}/v$ in Eq.\ (\ref{transmission_amplitude}) is of order 1. We estimate the temperature required for the zero temperature regime assumed above,  $kT \ll  |A_1| A$, to be of order $100 \rm mK$. Thus, the interesting, exponentially suppressed, regime is well within experimental reach.

{\it Experimental setup II.} Now we consider the opposite limit, namely a rectangular device with $L_{12}\gg L_{13}$ and $\mu_i=\mu_j,\;\forall i,j$. Without laser irradiation no net current flows between the terminals. Nevertheless, a large counterclockwise background current exists, as illustrated by the gray arrows in Fig. \ref{device_figure}. When the laser is switched on  parts of the large counterclockwise current are blocked. According to Eq. (\ref{transmission_amplitude}), the blocking will be more effective for the long edges than for the short edges and thus a net current from terminal 3 to terminal 1 and one from terminal 2 to terminal 4 will be measured.  We had suppressed this effect in the scenario of setup I by choosing  all edges of equal length. The action of the laser on the background current then cancels between edges. Alternatively, one may observe the suppression predicted by Eq.\ (\ref{I12}) also for unequal lengths $L_{12} \neq L_{13}$ by  measuring the response of the current $I_{12}$ to a small change of $\mu_1$.

In conclusion, we have shown that it is possible to optically manipulate the electronic transport in quantum hall edge states by illumination with properly tuned laser light. Remarkably, the backscattering into counterpropagating modes by photons is only possible if the relevant edge modes are hole-like. The scattering of hole-like edge states to electron-like edge states reverses the group velocity which results in a measurable reversal of the charge current direction through an edge. The HgTe QW is especially convenient for an experimental realization of this proposal, since the relevant parameters are highly tunable and well under control.

We acknowledge enlightening discussions with B. Braunecker, H. Buhmann, D. Loss and L.W. Molenkamp. M.J.S. was financially supported by Swiss NSF and NCCR Nanoscience. E.G.N. acknowledges financial support by the German DFG via grant no. AS327/2-1. B.T. was financially supported by the German DFG via grant no. Tr950/1-1.


\begin{thebibliography}{99}
\bibitem{e_h_like_footnote} The characterization of Landau levels as electron (hole)-like refers to their behavior near an edge: if the energy of a band increases (decreases) as the edge is approached, we label it as electron (hole)-like.
\end{thebibliography}
\end{document}